\documentstyle[12pt]{article}
\relax
\def\akpi{C_{11}<\pi^+\pi^-|O_{11}|K^0>}
\def\l6{\lambda_6}
\def\gsim{{~\raise.15em\hbox{$>$}\kern-.85em \lower.35em\hbox{$\sim$}~}}
\def\lsim{{~\raise.15em\hbox{$<$}\kern-.85em \lower.35em\hbox{$\sim$}~}}
\begin{document}
\begin{titlepage}
\vfill

\hskip 4in {ISU-HET-99-8}

\hskip 4in {September 1999}
\vspace{1 in}
\begin{center}
{\large \bf Comment on the Matrix Element of $O_{11}$}\\ 

\vspace{1 in}
{\bf  Xiao-Gang~He$^{(a,b)}$} 
{\bf  and G.~Valencia$^{(c)}$}\\
{\it  $^{(a)}$ Department of Physics,
               National Taiwan University,
               Taipei 10674}\\
{\it  $^{(b)}$ School of Physics,
               University of Melbourne,
               Parkville, Vic. 3052}\\
{\it  $^{(c)}$ Department of Physics and Astronomy,
               Iowa State University,
               Ames IA 50011}\\
\vspace{1 in}
     %	{\large \bf ABSTRACT}
\end{center}
\begin{abstract}

The gluon dipole operator, $O_{11}$,  
has received much attention recently because it can have 
a large coefficient in some SUSY extensions of the 
standard model. We find that the commonly 
used  matrix element of $O_{11}$ of Bertolini, Eeg and 
Fabbrichesi is in rough (but accidental) numerical agreement 
with an estimate based on dimensional analysis.  

\end{abstract}

\end{titlepage}

\clearpage

One of the operators in the low energy effective Hamiltonian 
responsible for $|\Delta S|=1$ weak decays is the gluon dipole 
operator, which can be written as ($H_{eff}=C_{11}O_{11}$), 
\begin{equation}
H_{eff} =  C_{11}{g_s \over 8 \pi^2}
\bar s \bigl[ m_d R + m_s L\bigr]  
T^a G_a^{\mu\nu} \sigma_{\mu\nu}~+~{\rm h.c.},
\label{sdgver}
\end{equation}
where $G^{\mu\nu}_a$ is the gluon field 
strength tensor, and $L,R \equiv (1\mp\gamma_5)/2$. 
In the standard model, the coefficient $C_{11}$ 
is sufficiently small to make the effects of this operator 
negligible. Beyond the standard model, 
however, this operator can have a large coefficient \cite{kagan} 
and it becomes important to estimate its matrix element. 
The operator has received much attention recently, in 
connection with $\epsilon^\prime/\epsilon$ \cite{recent}.

The importance of this operator for kaon decays, and in 
particular for the analysis of $\epsilon^\prime/\epsilon$ dates 
back to the Weinberg model of CP violation in the early 80's. 
It was shown back then, that one had to pay particular attention to 
the chiral properties of the operator \cite{donhol} in order to 
obtain physical amplitudes that obeyed the FKW theorem \cite{fkw}. 

The gluon-dipole operator transforms as $(\bar{3},3)$ or $(3,\bar{3})$ 
under chiral rotations and, in the standard model, it is proportional 
to the light quark masses. 
It is well known that there are no operators with these transformation 
properties in the lowest order, ${\cal O}(p^2)$, weak chiral 
Lagrangian \cite{wise,kambor}. At next to leading order, ${\cal O}(p^4)$, 
there are several operators with the desired properties. 

The strong interactions of pions and kaons are 
described to order ${\cal O}(p^4)$ in chiral perturbation 
theory  by the Lagrangian of Gasser and Leutwyler \cite{gl}. 
The ingredients to construct this Lagrangian are the non-linear function 
$\Sigma=\exp(2i\phi/f)$ which contains the octet of pseudo-Goldstone 
bosons $\phi$ and that transforms as $\Sigma \rightarrow R \Sigma L^\dagger$ 
under $SU_L(3)\times SU_R(3)$. Interactions that respect chiral 
symmetry are constructed in terms of derivatives of $\Sigma$. 
Explicit chiral symmetry 
breaking due to the non-zero light quark mass matrix $M=diag(m_u,m_d,m_s)$ 
is introduced through the factors $S$ and $P$,
\begin{eqnarray}
S &=& \chi^\dagger \Sigma + \Sigma^\dagger \chi \nonumber \\
P &=& i(\chi^\dagger \Sigma - \Sigma^\dagger \chi) 
\end{eqnarray}
For our present purpose it suffices to take $\chi = 2B_0 M$, where  
the parameter $B_0$ is proportional to the quark condensate $<\bar{q}q>$, 
and relates the current quark masses to the meson masses:
\begin{equation}
<\bar{q}q> = -f_\pi^2 B_0,\;\;\; m_K^2 = B_0(m_s+m_d),\;\;\;
m_\pi^2 = B_0(m_u+m_d)
\end{equation}

The leading order weak chiral Lagrangian transforming as $(8_L,1_R)$ 
under chiral symmetry was first written down 
by Cronin \cite{cronin} and has only one term 
(We use the notation $L_\mu = i\Sigma^\dagger D_\mu \Sigma$),
\begin{equation}
{\cal L}_W^{(2)}=c_2{\rm Tr}\biggl(\lambda_6 L^2 \biggr)
\label{lw2}
\end{equation}
To introduce explicit chiral symmetry breaking due to the non-zero 
light quark mass matrix $M$ into the chiral Lagrangian one pretends 
that the mass matrix transforms as $M \rightarrow R M L^\dagger$ and 
constructs operators with the desired transformation properties. 
For the dominant $(8_L,1_R)$ weak operators 
it is well known that there is no mass term at order $p^2$, 
in accordance with the FKW theorem. 

The operators that occur at next to leading order in the weak 
chiral Lagrangian have been written down in Ref.~\cite{kambor}. 
From their general list, those that could correspond to chiral realizations 
of the operator $O_{11}$ are the ones that contain one factor 
of the quark masses, they are (only five are independent) 
\cite{kambor}
\begin{eqnarray}
{\cal L}^{(4)}_W &=&
 E_{10} {\rm Tr}\biggl(\l6\{S,L^2\}\biggr)
+E_{11}{\rm Tr}\biggl(\l6 L_\mu S L^\mu\biggr)
+E_{12}{\rm Tr}\biggl(\l6L_\mu \biggr){\rm Tr}\biggl(\{L^\mu,S\}\biggr) 
\nonumber \\
&+& 
E_{13} {\rm Tr}\biggl(\l6 S\biggr){\rm Tr}\biggl(L^2\biggr)+
E_{14} {\rm Tr}\biggl(\l6 L^2\biggr){\rm Tr}\biggl(S\biggr)
+E_{15} {\rm Tr}\biggl(\l6i[P,L^2]\biggr)
\label{lw4} 
\end{eqnarray}
From the point of view of the transformation properties of $O_{11}$ 
under chiral symmetry all these terms are equally valid. 
It has been argued in Ref.~\cite{berto} that the precise form of the 
short distance operator in Eq.~\ref{sdgver} requires a chiral 
representation in which $\lambda_6$ and the quark mass matrix 
appear next to each other. This requirement reduces the possible 
operators in Eq.~\ref{lw4} to the ones multiplying $E_{13}$ and 
the combination $E_{10}-E_{15}$. 

Using Eq.~\ref{lw4} we find for the matrix element of $O_{11}$,
\begin{eqnarray}
\akpi &=&- {2\sqrt{2}\over f_\pi^3}
(m_K^2-m_\pi^2)\biggl( (m_K^2+2m_\pi^2)E_{10} +m_\pi^2 E_{11}
\nonumber \\
&&
+(4m_\pi^2-2m_K^2)E_{13}+(m_\pi^2+2m_K^2)E_{14}
+m_K^2 E_{15}\biggr)
\label{weakclme}
\end{eqnarray}
From this expression it is clear that the matrix element, in general, 
does not vanish in the limit $m_\pi\rightarrow 0$. 

Although the framework of Eq.~\ref{lw4} is completely general, 
it does not tell us the size of the coefficients $E_i$. We can 
estimate the size of the $E_{10}$, for example, that is needed 
to match the operator $O_{11}$ by using naive dimensional 
analysis \cite{wein}. For this purpose we write down a Lagrangian 
with the minimal number of fields that is contained in the term 
proportional to $E_{10}$ 
(other terms are related to this one by soft pion theorems), 
${\cal L} = g_M m_K^2 \partial_\mu K \partial^\mu \pi$ and 
match $g_M$ to $C_{11}$ following Weinberg \cite{wein}, with the 
result
\begin{equation}
\biggl(E_{10}\biggr)_{NDA} 
\sim {f_\pi^2 \over 4 \sqrt{2}} {C_{11}\over 8 \pi^2}
{g_s\over 4\pi}
\label{nda}
\end{equation}
From this we write
\begin{equation}
\biggl(\akpi\biggr)_{NDA} \sim 
(m_K^2-m_\pi^2){m_K^2 \over f_\pi}
{C_{11}\over 16 \pi^2} {g_s \over 4\pi}
\label{result}
\end{equation}
To evaluate this expression one would use a value of 
$g_s \sim \sqrt{4\pi}$. Note that an equivalent expression 
is obtained if one uses the term proportional to $E_{13}$ 
in Eq.~\ref{weakclme} instead.

\newpage

\noindent{\bf Bag Model Estimate}

\vspace{.2in}

An explicit calculation of 
$<\pi|O_{11}|K>$ within the MIT bag model \cite{donhag}, 
supplemented with a soft 
pion theorem, led to the estimate of Donoghue and Holstein \cite{donhol},
\begin{equation}
\biggl(\akpi\biggr)_{MIT} =  -C_{11}{g_s m_s\over 32 \pi^2}
{A_{K\pi}\over 2 f_\pi}{m_K^2 \over\Lambda^2}
\label{dono}
\end{equation}
Eq.~\ref{dono}, is a trivial rescaling of the actual calculation in 
Ref.~\cite{donhol}. The factor $A_{K\pi}$ is obtained numerically 
from the Bag model, $A_{K\pi} = 0.4~{\rm GeV}^3$, and   
the last factor in Eq.~\ref{dono} introduces 
the suppression required by the FKW theorem. In accordance with 
power counting they choose $\Lambda\sim 1$~GeV, corresponding to 
a matching of $O_{11}$ into an ${\cal O}(p^4)$ chiral Lagrangian 
such as the term that multiplies $E_{10}$ in Eq.~\ref{lw4}. 
Numerically, we find that 
\begin{equation}
\biggl(\akpi\biggr)_{MIT}  \approx  0.8\  
\biggl(\akpi\biggr)_{NDA}
\end{equation}

\vspace{.2in}

\noindent{\bf Chiral Quark Model Estimate}

\vspace{.2in}

More recently, Bertolini, Eeg and Fabbrichesi have used a chiral 
quark model supplemented with some matching conditions to 
estimate that \cite{berto},
\begin{equation}
\biggl(\akpi\biggr)_{CQM} = {\sqrt{2}\over f_\pi^3} 
(m_s-m_d)m_\pi^2 {C_{11}\over 16 \pi^2} 
\biggl(-{11\over 4} <\bar{q}q>_G \biggr)
\label{chiqm}
\end{equation}
There are two points that we want to stress about this expression. 
First, the last factor in Eq.~\ref{chiqm} is the 
model dependent quantity that arises from their chiral quark model, 
and that we take at face value.  
Second, the factor of $m_\pi^2$ arises from the requirement that the 
short distance operator $O_{11}$ match into an order $p^4$ weak 
chiral Lagrangian of the form \cite{berto}
\begin{equation}
{\cal L}\sim {\rm Tr}\biggl[ (\Sigma^\dagger M \l6+\l6 M \Sigma)
D^\mu\Sigma^\dagger D_\mu \Sigma\biggr]
\label{bermat}
\end{equation}
Comparing this with the general form, Eq.~\ref{lw4}, we see that 
the only term that is retained is that proportional to the 
operator whose coefficient is $E_{10}-E_{15}$. 
As mentioned above, the requirement 
that the quark mass matrix $M$ and $\lambda_6$ appear next to 
each other selects this term plus the one whose coefficient is $E_{13}$. 
This latter one is dropped in Ref.~\cite{berto} because 
all products of two traces are suppressed in their 
model. 

The specific numerical result of Ref.~\cite{berto}, 
however, is very similar to our dimensional analysis estimate. This 
happens because the 
$m_\pi^2/m_K^2$ suppression that occurs in the weak chiral Lagrangian 
operators that occur in their matching
is compensated by the large numerical coefficient, 11, 
in Eq.~\ref{chiqm}. Numerically,
\begin{equation}
\biggl(\akpi\biggr)_{CQM} \approx 1.4\ 
\biggl(\akpi\biggr)_{NDA} 
\label{comp}
\end{equation}
We see that all three estimates are numerically very 
similar and in agreement with each other within the uncertainty 
of each approach. However, it is clear that the numerical 
agreement with the chiral quark model result of Ref.~\cite{berto}
is accidental. 

\vspace{.2in}
 
\noindent{\bf Beyond the Standard Model}

\vspace{.2in}

In all the cases that we have discussed, we have considered the 
matrix element of the operator written as in Eq.~\ref{sdgver}. 
In this form it appears that the operator vanishes in the chiral 
limit being proportional to the light-quark masses. This was, in fact, 
an important ingredient in the matching to a corresponding chiral 
Lagrangian. However, in some models of interest, this is just an 
artifact of the normalization; the coefficient $C_{11}$ goes as 
$m_s^{-1}$. In this case we want to construct a low energy 
meson Lagrangian that transforms as $(\bar{3},3)$ or $(3,\bar{3})$ 
but that is not proportional to the light quark masses. 

It is possible to write a term without derivatives,
\begin{equation}
{\cal L} = {\rm Tr}\biggl(\Sigma^\dagger h + h^\dagger \Sigma\biggr)
\end{equation}
that does not contribute to $K\rightarrow \pi\pi$ amplitudes once  
tadpoles are properly subtracted. The leading order amplitude 
arises from a Lagrangian with two derivatives. In terms of the matrix $h$ 
whose only non-zero entry is $h_{23}=1$ we can write, for example, 
\begin{eqnarray}
{\cal L}& =& g_{N1} 
{\rm Tr}\biggl(h D_\mu \Sigma^\dagger D^\mu \Sigma \Sigma^\dagger\biggr)
~+~{\rm h.c.}
\nonumber \\
&+& g_{N2}{\rm Tr}\biggl(\Sigma^\dagger h + h^\dagger \Sigma\biggr)
{\rm Tr}\biggl(D_\mu \Sigma^\dagger D^\mu\Sigma\biggr)
\label{newcl}
\end{eqnarray}
The first term in Eq.~\ref{newcl} yields a matrix element proportional 
to $m_\pi^2$, but the second term does not. We find,
\begin{equation}
\akpi = g_{N2}{2\sqrt{2}\over f_\pi^3}
m_K^2 + {\cal O}(m_\pi^2/m_K^2)
\end{equation}

Although we cannot compute $g_{N2}$, we can again estimate it 
with naive dimensional analysis following Weinberg. Noting that 
the operator does not contain terms with only two fields, we find
\begin{equation}
g_{N2}\sim f_\pi^3 \biggl( {g_s m_s C_{11} \over 16 \pi^2} \biggr)
\end{equation}
resulting in 
\begin{equation}
\biggl(\akpi\biggr) \sim m_s m_K^2 {g_s C_{11}\over 8 \pi^2}
\end{equation}
This result is equivalent to Eq.~\ref{result}, differing only by 
factors of order one that cannot be accounted for with naive 
dimensional analysis.

\vspace{1in}

\noindent {\bf Acknowledgments} The work of G.V. was supported in
part by DOE under contract number DEFG0292ER40730. 
The work of X-G.H. was supported by NSC of R.O.C. under grant number
NSC88-2112-M-002-041 and Australian Research Council. 
G.V. thanks the theory group at SLAC for their hospitality and 
for partial support while this work was completed. We thank 
S.~Bertolini, J.~Eeg, M.~Fabbrichesi and H.~Murayama for useful 
comments.


\begin{thebibliography}{99}

\bibitem{kagan}{A.~Kagan, {\it Phys. Rev.} {\bf D51} (1995) 6196.}

\bibitem{recent}
{D. Chang, X.-G. He and B. McKellar, hep-ph/9909357;
G.~Eyal, {\it et. al.}, hep-ph/9908382; 
A.~Buras, {\it et. al.}, hep-ph/9908371; 
B.~Holdom, hep-ph/9907361; 
S.~Baek et al., hep-ph/9907472;
A.~Masiero and H. Murayaman, Phys. Rev. Lett. {\bf 83}, 22(1999).
}

\bibitem{donhol}{J.~Donoghue and B.~Holstein, {\it Phys. Rev.} 
{\bf D32} (1985) 1152.}

\bibitem{fkw}{G.~Feinberg, P.~Kabir and S.~Weinberg, {\it Phys. Rev. Lett.} 
{\bf 3} (1959) 527.}

\bibitem{wise}{J.~Bijnens and M.~Wise, {\it Phys. Lett.}
{\bf 137B} (1984) 245.}

\bibitem{kambor}{J.~Kambor, J.~Missimer and D.~Wyler, {\it Nucl. Phys.} 
{\bf B346} (1990) 17.}

\bibitem{gl}{J.~Gasser and H.~Leutwyler, {\it Nucl. Phys.} {\bf B250} 
(1985) 465.}

\bibitem{cronin}{J.~A.~Cronin, {\it Phys. Rev.} {\bf 161} 
(1967) 1483.}

\bibitem{wein}{H.~Georgi and L.~Randall, {\it Nucl. Phys.} 
{\bf B276} (1986) 241; 
S.~Weinberg, {\it Phys. Rev. Lett.} {\bf 63} (1989)
2333.}

\bibitem{donhag}{J.~Donoghue, J.~Hagelin and B.~Holstein, 
{\it Phys. Rev.} {\bf D25} (1982) 195.}

\bibitem{berto}{S.~Bertolini, J.~Eeg and M.~Fabbrichesi, 
{\it Nucl. Phys.} {\bf B449} (1995) 197.}

\end{thebibliography}
\end{document}